\documentclass[aps,superscriptaddress]{revtex4}

\newcommand{\M}{{\cal M}}
\newcommand{\T}{{\cal T}}
\newcommand{\E}{{\cal E}}
\newcommand{\C}{{\cal C}}
\newcommand{\V}{{\cal V}}
\newcommand{\CP}{{\cal C}^\perp}
\newcommand{\VP}{{\cal V}^\perp}

\begin{document}


\begin{abstract}
\vspace*{1cm}
\noindent A general class of authentication schemes for arbitrary quantum
messages is proposed. The class is based on the use of sets of unitary
quantum operations in both transmission and reception, and on appending
a quantum tag to the quantum message used in transmission. The previous
secret between partners required for any authentication is a classical
key. We obtain the minimal requirements on the unitary operations that
lead to a probability of failure of the scheme less than one. This failure
may be caused by someone performing a unitary operation on the message
in the channel between the communicating partners, or by a potential
forger impersonating the transmitter.  
\end{abstract}

\title{Quantum authentication with unitary coding sets}

\author{Esther P\'erez}
\affiliation{ETSIT, Universidad de Vigo, Campus Universitario s/n,
E-36200 Vigo, Spain}

\author{Marcos Curty}
\affiliation{ETSIT, Universidad de Vigo, Campus
Universitario s/n, E-36200 Vigo, Spain}
\affiliation{Quantum Information Theory Group, ZEMO (Zentrum f\"ur
Moderne Optik), University Erlangen-N\"urnberg, Staudtstr. 7/B2,
D-91058 Erlangen, Germany}

\author{David J. Santos}
\email[E-mail address: ]{dsantos@com.uvigo.es}
\affiliation{ETSIT, Universidad de Vigo, Campus Universitario s/n,
E-36200 Vigo, Spain}

\author{Priscila Garc\'{\i}a-Fern\'andez}
\affiliation{Instituto de \'Optica, CSIC, Serrano, 123, E-28006
Madrid, Spain}

\maketitle

\section{Introduction}

Providing a way to check the integrity of information transmitted
over or stored in an unreliable medium is of prime concern to 
the fields of open computing and communications. Mechanisms that
provide such integrity check are called message authentication
schemes \cite{MENEZES_1996}. They were originally proposed by
Gilbert and co-workers \cite{GILBERT_1974}, while the general
theory of unconditional authentication was developed by Simmons
(see e.g. \cite{SIMMONS_1992}).

In the analysis of any message authentication scheme one has to
consider three participants: A transmitter (Alice), a receiver (Bob),
and an opponent (Eve).  Alice wishes to communicate some information
(the plain-text) to Bob using a public communications channel; Bob, in
turn, would like to be confident that any information received actually
came from Alice, rather than from some third party (Eve). Much as in the
usual encryption scenario, classical cryptography provides two different
approaches to authentication: secret-key and public-key authentication. In
this paper we will focus on the first scenario, the secret-key setting,
and therefore we shall assume that Alice and Bob share some secret
key previously established in a secure manner. This key allows Alice
to select an encoding rule (a one-to-one function between the set of
plain-texts and the set of messages, also called cipher-texts), chosen
from a predetermined set, and encode the plain-text to obtain the message,
which is then sent to Bob through the channel.  The encoding rule, which
is usually changed every time a new message is transmitted, defines a
set of valid messages.  Upon receiving a message, Bob accepts it as
being authentic (i.e. as coming from Alice) if and only if it belongs
to that set, in which case he will recover the plain-text applying the
corresponding decoding rule. This decoding rule is well-defined since
each encoding rule is one-to-one.

A special class of secret-key message authentication schemes are
message authentication codes (MACs), which contain the plain-text in
the clear. In this case, each encoding rule generates, depending on
the value of the actual plain-text, an authenticator or tag that is
appended to the plain-text before actually sending it. MACs decoding
rules return, depending on the plain-text and the tag, a bit indicating
when Bob must regard the message as authentic, and accepting it as coming
from Alice, and when he must discard it. The basic requirement is that
the tags, which are produced by the encoding rule, must be accepted as
valid when the matching decoding rule is used on reception.  When an
authentication protocol fulfills this requisite it is said to provide
perfect deterministic decoding.  Wegman and Carter \cite{WEGMAN_1981}
introduced several information-theoretic secure constructions for such
schemes.  Basically, their techniques use universal hash families as the
coding set. To generate the tag, Alice uses a particular hash function,
selected from the universal hash family by the secret key. This action
compresses the plain-text to a smaller string of bits. The string of bits
can be later encrypted using the Vernam cipher \cite{VERNAM_1926}. This
last step allows to re-use the encoding rules, since Eve cannot obtain
any information about the particular hash function used by Alice and Bob.

The possibility of employing quantum resources to obtain more efficient
classical-message quantum authentication schemes is still an open
issue. In \cite{CURTY_2001} the authors showed that, using quantum
effects, the authentication of a binary classical message is possible with
a key of length shorter than the one required by classical schemes. However,
it is not clear yet whether more efficient quantum techniques exist for
longer messages.

While the authentication of classical messages is a fundamental
topic in classical communications networks, the analogous quantum
problem, the authentication of quantum states, could also become
important in future quantum information communication systems.
Leung has addressed this question partially in \cite{LEUNG_2002}. Her
proposal is based on a modification of the private quantum channel
\cite{AMBAINIS_2000,BOYKIN_2000}. The authentication process requires
a classical secret key, a quantum communications channel, and an
authenticated two-way classical one. Another classical
secret-key quantum authentication protocol, but without additional classical
communication, has been proposed in \cite{BARNUM_2002}. This scheme
uses stabilizer purity testing codes, and its error probability
decreases exponentially with the length of the quantum tag. Rather
surprisingly, the authors also showed that any protocol that guarantees
unconditionally secure authenticity must encrypt the quantum plain-text
almost perfectly.  This fact contrasts with classical MACs, where
encryption of the plain-text is not necessary for unconditional
security. More recently, Gea-Banacloche \cite{GEA_BANACLOCHE_2002}
has approached data authentication from an steganographic perspective,
making use of standard quantum correcting code techniques.

In this paper we analyse a special class of authentication schemes for
quantum messages. In particular, we study those which use classical secret
keys and coding sets composed of unitary operations. Our main goal is
to find the minimal conditions that must be satisfied by any unitary
coding set so as to make quantum authentication possible.  We analyse
the security of the scheme under two general attacks. In the first one,
the unitary attack, Eve, who knows both the coding and decoding sets,
tries to modify the cipher-text by means of a unitary operation. In the
second, the forgery attack, Alice has not initiated the transmission yet
(or it may have been blocked by Eve), and Eve attempts to prepare a new
fake quantum message with the purpose of passing Bob's verification test.

The paper is organized as follows. In Sec.\ II we describe general
authentication schemes for quantum messages, and we introduce those
which use unitary coding sets. We also present some notation used in the
paper. Sec.\ III analyzes the unitary attack, and Sec.\ IV focuses on
the forgery attack. In Sec. V we discuss the restrictions that these
two attacks impose upon both coding sets.  Finally, we present our
conclusions in Sec.\ VI.

\section{Authentication of quantum messages}

Suppose Alice needs to send a certified quantum message to Bob.
Her goal is to make Bob confident about the authenticity of the message
and sender. If we consider an scenario where both participants share a
quantum secret key (for example, a set of EPR pairs), and they have access
to an authenticated classical channel, then the solution is quite simple:
Alice can just use quantum teleportation \cite{BENNETT_1993} to send
the quantum plain-text. However, the reliable storage and manipulation
of entangled quantum objects is not technologically available yet, so
a more practical situation arises when the secret key shared by the two
partners is classical.

Analogously to the classical setting, quantum authentication with
a classical secret key can be performed in three phases:

\begin{enumerate}

\item \textit{Tagging}: To certify her message (the plain-text), Alice
appends to it a particular public-known quantum state that we shall
call, following the standard classical notation, a tag.  Specifically,
we will assume that both message and tag quantum states belong,
respectively, to the state spaces $\M$ (${\rm dim}(\M)=M$), and $\T$
(${\rm dim}(\T)=T$). The space of tagged messages is defined, therefore,
as $\E=\M\otimes \T$. Alice and Bob also openly agree in a particular
splitting of the tag space $\T$ in the direct sum of two subspaces
$\T=\V \oplus \VP$, where $\V$ is considered the subspace of valid tags
and $\VP$ the subspace of invalid tags. This splitting of $\T$ leads
naturally to the direct sum $\E=\C\oplus\CP$, where $\C=\M\otimes \V$
will be the subspace of valid messages (usually called the code space;
${\rm dim}(\C)=C$), and $\CP=\M\otimes \VP$ will be the subspace of
invalid messages (${\rm dim}(\CP)=D$). The original tagged message has
the general form $\rho_\E= \rho_\M \otimes \rho_\T$, with $\rho_\M$
the plain-text quantum message and $\rho_\T$ any state in $\V$.

\item \textit{Encoding}: After the tagging procedure has been completed,
Alice, depending on the value of the $n$-bit key, $k$, shared with Bob,
performs an encoding rule on the tagged message.  In principle, encoding
(and decoding) rules could be trace-preserving completely positive maps
(TPCP), but in this paper we will restrict ourselves to the case of
unitary rules. Therefore, the encoding rule, $U(k)$, is selected from
the unitary coding set $\{U(0),\cdots,U(K-1)\}$, where $K=2^n$ and,
without loss of generality, we set $U(0)=I$. If the tagged message is
$\rho_\E$, then the state of the message Alice sends to Bob is given by

\begin{equation}
\rho_\E(k)=U(k)\rho_\E U^\dagger(k).
\end{equation}

\noindent Thus, a tagged message encoded by Alice with $U(k)$ will
necessarily belong to $\C_k$, the subspace of all the tagged messages
transformed by $U(k)$.

\item \textit{Verification}: To verify the authenticity of the received
message, Bob needs to check whether it belongs to $\C_{k}$ or not. In
the latter case, he should regard the message as invalid and so
discard it. One way in which Bob can make this check is performing
the matching decoding rule $U^\dagger(k)$ (from the decoding set
$\{U^\dagger(0),\cdots,U^\dagger(K-1)\}$) on the received encoded 
tagged message, and then measure the tag portion to see whether
the resulting tagged message belongs to $\C$ or to $\CP$. If it belongs
to $\C$, and no forgery on the encoded tagged message has taken place
in transit, he could unambiguously recover the original plain-text,
just tracing out the tag system.  
\end{enumerate}

In the next two sections we shall analyse the minimal conditions that
any unitary coding set must satisfy in order to make authentication
possible. It should be clear from the outset that we are not looking for
the optimal coding sets, defined as those which make the probability of
failure of the protocol minimal. Therefore, our scheme will be secure if
Eve can only break it in a probabilistic way. We shall study two general
attacks. In the unitary attack we shall regard Eve capable of modifying
the state in the channel by means of a unitary operation; in the case
of forgery we shall assume that Eve can intercept the state travelling
from Alice to Bob, discard it, and forge a new tagged message. In both
cases we shall frequently use operators acting on the space $\E$, and we
shall be interested in describing how these operators act on the $\C$
and $\CP$ subspaces. If we define the orthogonal projection operators
$P_i$ and $P_o$ as the ones that, respectively, project a state from $\E$
into $\C$ or $\CP$, an arbitrary operator $A_\E$ can be written as

\begin{equation}
A_\E= A_{ii}+ A_{io}+ A_{oi}+ A_{oo},
\label{DESCOMPOSICION_IO}
\end{equation}

\noindent where $A_{jk}=P_jA_\E P_k$, with $j,k=i,o$. If the decomposition
(\ref{DESCOMPOSICION_IO}) is used in operator expressions of the form
$\chi_\E=A_\E \rho_\E A_\E^\dagger$, the corresponding `i-o' operators
are related by the matrix equation

\begin{equation}
\left(
\begin{array}{cc}
\chi_{ii} & \chi_{io}\\
\chi_{oi} & \chi_{oo}\\
\end{array}
\right)
=
\left(
\begin{array}{cc}
A_{ii} & A_{io}\\
A_{oi} & A_{oo}\\
\end{array}
\right)
\left(
\begin{array}{cc}
\rho_{ii} & \rho_{io}\\
\rho_{oi} & \rho_{oo}\\
\end{array}
\right)
\left(
\begin{array}{cc}
A_{ii}^\dagger & A_{oi}^\dagger\\
A_{io}^\dagger & A_{oo}^\dagger\\
\end{array}
\right),
\end{equation}

\noindent where $A_{jk}^\dagger=P_k A_\E^\dagger P_j$.

\section{The unitary attack}
\label{ATAQUE_UNITARIO} 

Let us assume that Eve performs a unitary quantum operation $F_\E$
on the encoded tagged message in transit between Alice and Bob. This
operation changes the state of the encoded message from $\rho_\E(k)$
to $F_\E \rho_\E(k) F_\E^\dagger$.  Bob, ignorant about this malicious
action, will perform his decoding operation on the encoded tagged message
received, obtaining as the decoded tagged message the state

\begin{equation}
\rho_\E^E(k)=U^\dagger(k)F_\E \rho_\E(k) F_\E^\dagger 
U(k)=Q(k) \rho_\E Q^\dagger(k),
\label{NUEVA_2}
\end{equation}

\noindent where $Q(k)=U^\dagger(k) F_\E U(k)$. Thus,
Eve will be successful in her attack if, when
the $k$ key was used, $\rho_\E^E(k) \in \C$.
This condition can also be written in terms of the action of the $P_i$
operator as

\begin{equation}
P_i \rho_\E^E(k) P_i=\rho_\E^E(k).
\label{NUEVA_1}
\end{equation}

\noindent Since Eve does not know either which key Alice and Bob used
or the actual message sent, $\rho_\E$, the probability of her action
being unnoticed is one if and only if

\begin{equation}
\rho_\E^E(k)\in \C, \quad \forall k,\quad \forall \rho_\E \in \C.
\label{CONDICION_PROBABILIDAD_UNO}
\end{equation}

\noindent Using Eq. (\ref{NUEVA_2}) in Eq. (\ref{NUEVA_1}), one can
rewrite (\ref{CONDICION_PROBABILIDAD_UNO}) as

\begin{equation}
Q_{oi}(k)=0, \quad \forall k.
\label{CONDICION_EXITO_F}
\end{equation}

\noindent Given the unitarity of the $Q(k)$ operators, Eq.
(\ref{CONDICION_EXITO_F}) implies that $Q_{io}(k)=0, \quad \forall
k$. In particular, for $k=0$, we have that $Q_{oi}(0)=Q_{io}(0)=0$,
which, since $U(0)=I$, requires that $F_{oi}=F_{io}=0$. All these
restrictions on $F_\E$ can be summarized in the following commutators:

\begin{eqnarray}
[F_\E,P_i]&=&0, \label{CONDICION_ALFA}\\
{[F_\E,P_i(k)]}&=&0, \quad \forall k\neq 0,
\label{CONDICION_BETA}
\end{eqnarray}

\noindent where the $P_i(k) = U(k) P_i U^\dagger(k)$ are the projectors
resulting from the transformation of the $P_i$ by the unitary operations
$U(k)$. The commutator (\ref{CONDICION_ALFA}) requires Eve to use a
block-diagonal unitary operator in her attack. The block structure of
$F_\E$ depends on how Alice and Bob decided to split $\E$ into $\C$ and
$\CP$. This information is public, so we shall assume that Eve can always
fulfill (\ref{CONDICION_ALFA}). But, can she always fulfill the $K-1$
commutators in (\ref{CONDICION_BETA})?  The $P_i(k)$ in these
commutators can be written, making use of the `i-o' decomposition, as

\begin{equation}
\left(
\begin{array}{cc}
U_{ii}(k) U_{ii}^\dagger(k) & U_{ii}(k) U_{oi}^\dagger(k) \\
U_{oi}(k) U_{ii}^\dagger(k) & U_{oi}(k) U_{oi}^\dagger(k) \\
\end{array}
\right)
\equiv
\left(
\begin{array}{cc}
G_{ii}(k) & H(k) \\
H^\dagger(k)& G_{oi}(k) \\
\end{array}
\right).
\label{OPEREADORES_GH}
\end{equation}

\noindent Assuming the block-diagonal form for $F_\E$, and using 
the notation introduced in (\ref{OPEREADORES_GH}),
the commutators in (\ref{CONDICION_BETA}) can be expressed in block form
as:

\begin{eqnarray}
[F_{ii},G_{ii}(k)]&=&0,\label{CONDICION_A}\\
{[F_{oo},G_{oi}(k)]}&=&0,\label{CONDICION_B}\\
F_{ii}H(k)&=&H(k)F_{oo},\label{CONDICION_C}\\
F_{oo}H^\dagger(k)&=&H^\dagger(k)F_{ii}, \label{CONDICION_D}
\end{eqnarray}

\noindent $\forall k\neq 0$. Therefore, Eve would be always successful
in her attack if she could find two unitary operators
$F_{ii}$ and $F_{oo}$ obeying (\ref{CONDICION_A})--(\ref{CONDICION_D}).

\subsection{A simpler case: $K=2$}

The problem posed by Eqs.
(\ref{CONDICION_A})--(\ref{CONDICION_D}) is quite complex. In order to
gain some insight into its possible solution, we shall begin studying the
simpler case $K=2$ (Alice and Bob share a key of just one bit). In this
case, Eqs. (\ref{CONDICION_A})--(\ref{CONDICION_D}) reduce to

\begin{eqnarray}
[F_{ii},G_{ii}]&=&0,\label{CONDICION_A_SIMPLE}\\
{[F_{oo},G_{oi}]}&=&0, \label{CONDICION_B_SIMPLE}\\
F_{ii}H&=&HF_{oo}, \label{CONDICION_C_SIMPLE}\\
F_{oo}H^\dagger&=&H^\dagger F_{ii}. \label{CONDICION_D_SIMPLE}
\end{eqnarray}

\noindent Since $G_{ii}$ and $G_{oi}$ are Hermitian operators, and because
one can always find an operator that commutes with an Hermitian
operator \cite{HORN_1985}, we could begin solving the equations above
selecting the $F_{ii}$ that fulfills (\ref{CONDICION_A_SIMPLE})
or, alternatively, selecting the $F_{oo}$ that fulfills
(\ref{CONDICION_B_SIMPLE}). Then we could obtain $F_{oo}$ (or $F_{ii}$)
from (\ref{CONDICION_C_SIMPLE}).  This last step, however, requires a
previous discussion about the nonsingular character of $H$. First, note
that $H$ need not be square (it would be square if dim($\C$)=dim($\CP$);
i.e. if $C=D$); also note that we do not know the rank of $H$. To
``invert'' $H$ in these circumstances we have to apply
the Singular-Value Decomposition Theorem for general matrices
\cite{HORN_1985}. Let  us consider separately the cases of maximum and
non maximum rank.

\subsubsection{The rank of $H$ is maximum} 

If $C=D$, then ${\rm rank}(H)=C$, ${\rm det}(H)\neq 0$, and therefore $H$
is nonsingular.  If $C\neq D$, we can use the Singular-Value Decomposition
of $H$ to show that:

\begin{enumerate}
\item If $C<D$, there exists an operator $J$ such that $HJ=I_{C\times C}$.
\item If $C>D$, there exists an operator $J$ such that $JH=I_{D\times D}$.  
\end{enumerate}

\noindent In both cases it can be shown that Eqs.
(\ref{CONDICION_A_SIMPLE})--(\ref{CONDICION_D_SIMPLE}) have a non-trivial
solution for $F_{ii}$ and $F_{oo}$. For example, if $C<D$ we could select
a non-trivial $F_{oo}$ obeying (\ref{CONDICION_B_SIMPLE}), then $F_{ii}$
would be uniquely obtained from (\ref{CONDICION_C_SIMPLE}), and it is
not difficult to show that they would also obey (\ref{CONDICION_A_SIMPLE})
and (\ref{CONDICION_D_SIMPLE}).

\subsubsection{The rank of $H$ is not maximum} 

Suppose ${\rm rank}(H)=N<C\leq D$. The Singular-Value Decomposition leads
in this case to $H=V \Sigma W^\dagger$, where $V$ and $W$ are unitary
matrices, and $\Sigma$ is the diagonal matrix containing in its diagonal
the singular values of $H$. Note that, since ${\rm rank}(H)<C$, part of
the singular values will be zero. Therefore, it is straightforward to
see that one can find in this case a matrix $J$ such that

\begin{equation}
HJ= V
\left(
\begin{array}{c|c}
I_{N\times N} & 0\\ \hline
0 & 0\\
\end{array}
\right)
V^\dagger.
\end{equation}

\noindent This would be the ``inversibility'' condition
for $H$ in this case. Although the algebra is a little bit
more complicated now, also in this case one can show that Eqs.
(\ref{CONDICION_A_SIMPLE})--(\ref{CONDICION_D_SIMPLE}) have a non-trivial
solution for $F_{ii}$ and $F_{oo}$. The same conclusion can be reached
if ${\rm rank}(H)=N$ and $C>D$. Note, however, that now the family of
solutions is bigger than in the former case, because, given $F_{oo}$,
there is more than one $F_{ii}$ fulfilling (\ref{CONDICION_C_SIMPLE}).

In summary, we have shown that when $K=2$ (Alice's and Bob's keys are
one-bit long), Eve can always successfully manipulate Alice's message
and pass Bob's verification test. Therefore, one-bit keys, independently
of the length of the tag appended to the message and of the length of the
message itself, cause the failure of the protocol. This result generalizes
our conclusion in \cite{CURTY_2002} to arbitrary message and tag spaces.

\subsection{$K>2$}

In the preceding Subsection we have seen that Eve can select
a unitary quantum operation, $F_\E$, such that conditions
(\ref{CONDICION_A})--(\ref{CONDICION_D}) are satisfied for
$k\leq 1$. Let us see whether she can select a $F_\E$ obeying also
(\ref{CONDICION_A})--(\ref{CONDICION_D}) $\forall k>1$.  Consider the
first of the conditions, Eq. (\ref{CONDICION_A}).  This equation, for the
case of the preceding Subsection ($k\leq 1$), is $[F_{ii},G_{ii}(1)]=0.$
Since $G_{ii}(1)$ is Hermitian, it can be written in diagonal form:

\begin{equation}
G_{ii}(1)=
\left(
\begin{array}{ccc}
\lambda_{ii}^1(1) & {} & { \bf 0}\\
{} & \ddots & {} \\
{ \bf 0} & {} & \lambda_{ii}^C(1)
\end{array}
\right),
\end{equation}

\noindent where the $\lambda_{ii}^r(1)$ ($r=1,\cdots,C$) are the
eigenvalues of $G_{ii}(1)$. If we denote the $p,q$ element of $F_{ii}$ as
$a_{ii}(p,q)$, the condition $[F_{ii},G_{ii}(1)]=0$ requires that 

\begin{equation}
\lambda_{ii}^p(1)=\lambda_{ii}^q(1),\quad \forall p \neq q;\,
p,q=1,\cdots,C,
\label{CONDICION_1}
\end{equation}

\noindent or 

\begin{equation}
a_{ii}(p,q)=0, \quad \forall p \neq q;\, p,q=1,\cdots,C.
\label{CONDICION_2}
\end{equation}

\noindent Eq. (\ref{CONDICION_1}) is a requirement on Alice and Bob,
so they, in order to protect themselves from Eve's attack, can by
design avoid it, just selecting $G_{ii}(1)$ with all its eigenvalues
different. This can be done only in the case that $C\leq D$, because if
$C>D$ the unitarity  of $U(1)$ shows that at least $C-D$ eigenvalues
have to be equal to one. Thus, from now on we will impose $C \leq D$ and
consider separately the cases $C=D$ and $C<D$. In both cases, however,
Eve will have to choose $F_\E$ obeying (\ref{CONDICION_2}), i.e. $F_\E$
must be diagonal in the same base in which $G_{ii}(1)$ is diagonal:

\begin{equation}
F_{ii}=
\left(
\begin{array}{ccc}
a_{ii}^1 & {} & { \bf 0}\\
{} & \ddots & {} \\
{ \bf 0} & {} & a_{ii}^C
\end{array}
\right),
\label{FORMA_DE_FII}
\end{equation}

\noindent where $|a_{ii}^r|=1$ ($r=1,\cdots,C$).

\subsubsection{The case $C=D$} 

Let us see if the diagonal form of $F_{ii}$ is
compatible with, for instance, the condition $[F_{ii},G_{ii}(2)]=0$. If
we denote by $g_{ii}^{(2)}(p,q)$ the $p,q$ element of $G_{ii}(2)$,
this commutator is zero when

\begin{equation}
a_{ii}^p=a_{ii}^q, \quad \forall p\neq q,
\label{CONDICION_1_BIS}
\end{equation}

\noindent or

\begin{equation}
g_{ii}^{(2)}(p,q)=0, \quad \forall p\neq q.
\label{CONDICION_2_BIS}
\end{equation}

\noindent Again, Eq. (\ref{CONDICION_2_BIS}) is a requirement
on Alice and Bob, and they can, by design, avoid it. This means
that Eve, in order to be successful in her attack, would have to
fulfill (\ref{CONDICION_1_BIS}). But this condition implies that
$F_{ii}=\exp(i\alpha)I_{C\times C}$, with $\alpha$ any angle. Since $C=D$
and ${\rm rank}[H(1)]$ is maximum (which depends only on Alice and
Bob), $F_{oo}$ is uniquely determined by (\ref{CONDICION_C_SIMPLE})
and must also have the form $F_{oo}=\exp(i\alpha)I_{C\times C}$.
In conclusion, $F_\E$ would be a module-one multiple of the identity
operation, but this is the trivial solution (Eve does not modify the
state in the channel). 

In summary, we have shown that if Alice and Bob select
the splitting of $\E=\C \oplus \CP$,  $U(1)$ and $U(2)$ such that

\begin{enumerate}
\item ${\rm dim}(\C)= {\rm dim}(\CP)$
\item  ${\rm rank}[H(1)]$ and  ${\rm rank}[H(2)]$ are maximum,
\item $G_{ii}(1)$ and $G_{ii}(2)$  have all their eigenvalues different,
\item and $G_{ii}(1)$ and $G_{ii}(2)$ do not share any eigenvector, 
\end{enumerate}

\noindent then Eve cannot, performing a unitary operation on the state
in the channel, produce the failure of the protocol in a deterministic
way. Thus two bits of key are sufficient to make the probability of Eve
being unnoticed less than one.

The four conditions above can be analysed from a geometric
perspective. The first one says that half of the tag space is considered
valid by Alice and Bob, which in fact is equivalent to having just one
qubit of tag. In the second condition $H(1)=U_{ii}(1)U_{oi}^\dagger(1)$,
so its rank is maximum and equal to $C$ only if the rank of $U_{ii}(1)$
and $U_{oi}(1)$ are also $C$. This means that $\C_1$, the subspace of
valid messages transformed by $U(1)$, is such that no state in $\C_1$
has null projection over $\C$ nor $\CP$.  In other words, $\C_1$
is maximally spread over the original valid and invalid message
subspaces, and the same holds for $U(2)$. In the third condition,
$G_{ii}(1)=U_{ii}(1)U_{ii}^\dagger(1)$, thus the basis where $G_{ii}(1)$
is diagonal represents a basis of $\C$ with the following property: its
image under $U^\dagger(1)$, projected over $\C$, is a $C$-dimensional
set of orthogonal vectors, and the norm of each projected vector is
given by the corresponding eigenvalue of $G_{ii}(1)$. If the eigenvalues
are all different then this basis is unique up to some arbitrary global
factors. Finally, the fourth condition says that the above bases for
$G_{ii}(1)$ and $G_{ii}(2)$ are maximally spread one over the other.

\subsubsection{The case $C<D$} 

If $C<D$ $F_{oo}$ is not uniquely determined by (\ref{CONDICION_C_SIMPLE})
and the situation is more complex.  From now on we shall assume that
$D$ is a multiple integer of $C$, so $D=qC$, $q>1$. This is in fact the
natural situation when using qubits, since both $D$ and $C$ are powers
of two.  Following the argument of the preceding section, 
in the bases of $\C$ and $\CP$ where $G_{ii}(1)$ and $G_{oi}(1)$ are
respectively diagonal, from (\ref{CONDICION_C}), with $k=1$, we get:

\begin{equation}
F_{oo}= 
\left(
\begin{array}{c|c}
\exp(i\alpha) I_{C} & 0\\ \hline
0 & W_1\\
\end{array}
\right),
\end{equation}

\noindent with $W_1$ an arbitrary unitary operator in a $(q-1)C$
dimensional space. But we may obtain further restrictions on $W_1$
from (\ref{CONDICION_B})--(\ref{CONDICION_D}).  First, note that
$[F_{oo},G_{oi}(1)]=0$ for all $W_1$. The reason is that, although
$G_{oi}(1)$ is diagonal, $W_1$ does not have to since $G_{oi}(1)$ has at
least $(q-1)C$ zero eigenvalues (recall that it is a $qC$-dimensional
hermitian operator and its maximum rank is $C$).  Let us write the matrix
representation of $G_{oi}(1)$ as

\begin{equation}
G_{oi}= 
\left(
\begin{array}{c|c}
\Lambda_{oi}(1) & 0\\ \hline
0 & O_{(q-1)C} \\
\end{array}
\right),
\end{equation}

\noindent where $\Lambda_{oi}(1)$ is the diagonal $C$-dimensional matrix
containing the non-zero eigenvalues of $G_{oi}(1)$, and $O_{(q-1)C}$ is
the zero, $(q-1)C$-dimensional matrix.  Now Alice and Bob, in order to
choose a $U(2)$ such that $G_{oi}(1)$ and $G_{oi}(2)$ do not share any
eigenvector, would like to restrict $W_1$ as much as possible.  But,
since both operators have $(q-1)C$ zero eigenvalues (i.e. ${\rm dim}
\left( {\rm ker}[G_{oi}(1)] \right)= {\rm dim} \left( {\rm ker}[G_{oi}(2)]
\right) = (q-1)C$), this cannot be done. The reason is that the total
space where $G_{oi}(k)$ acts has dimension $qC$, so  ${\rm dim} \left\{
{\rm ker}[G_{oi}(1)] \bigcap  {\rm ker}[G_{oi}(2)] \right\} \geq (q-2)C$.
Thus, $G_{oi}(1)$ and $G_{oi}(2)$ share at least $(q-2)C$ eigenvectors.

For simplicity, although this might not be the optimal situation, let us
further assume that Alice and Bob choose $U(2)$ such that $G_{oi}(1)$
and $G_{oi}(2)$ are diagonal in the same basis, but $G_{oi}(2)$ has
its $C$ non-zero eigenvalues shifted in the following way:

\begin{equation}
G_{oi}(2)= 
\left(
\begin{array}{c|c|c}
O_{C} &0 & 0\\ \hline
0 & \Lambda_{oi}(2) &0 \\
\hline
0 & 0& O_{(q-2)C} \\
\end{array}
\right).
\end{equation}

\noindent If the eigenvalues of $G_{ii}(2)$ are all different, the
unitarity of $U(2)$ makes the eigenvalues of $G_{oi}(2)$ to be also
different (in fact, each pair sums up to 1). Then $[F_{oo}, G_{oi}(2)]=0$
if and only if $W_1$ has the following block diagonal form:

\begin{equation}
W_1= 
\left(
\begin{array}{c|c}
\Omega_{C} &0 \\ \hline
0 & W_2  \\
\end{array}
\right),
\end{equation}

\noindent with $\Omega_C$ a unitary, diagonal, $C$-dimensional matrix,
and $W_2$ any unitary $(q-2)C$-dimensional operator. Using the result
above, together with $F_{ii}= \exp(i\alpha) I_C$, in (\ref{CONDICION_C})
or (\ref{CONDICION_D}) with $k=2$, gives $\Omega_C= \exp(i\alpha)
I_C$, so we have:

\begin{equation}
F_{oo}= 
\left(
\begin{array}{c|c|c}
\exp(i\alpha) I_C &0 & 0\\ \hline
0 & \exp(i\alpha) I_C &0 \\
\hline
0 & 0& W_2 \\
\end{array}
\right).
\end{equation}

\noindent Following the same line of reasoning, it is not difficult to
see that, in order to make $F_{oo}=\exp(i\alpha) I_{qC}$, Alice and Bob
need $q+1$ unitary encoding operators, $U(0), \cdots ,U(q)$ such that:

\begin{enumerate}
\item  ${\rm rank}[H(k)]$ is maximum $\forall k>0$;
\item $G_{ii}(k)$  has all its eigenvalues different $\forall k>0$;
\item for at least two values of $k$, say $r,s \neq 0$, $G_{ii}(r)$
and $G_{ii}(s)$ do not share any eigenvector;
\item and $\forall k \neq k'$, $k, k' \neq 0$, the range of $G_{oi}(k)$
and $G_{oi}(k')$ span disjoint, orthogonal, $C$-dimensional subspaces
of $\CP$.  
\end{enumerate}

In the particular case in which we have messages of $m$ qubits, a tag of $t$
qubits, and just a one-dimensional valid tag subspace, $C=2^m$
and $D=(2^t-1)C$, so $q+1=2^t$, and the number of bits of the classical
key equals the number of tag qubits.

\section{The forgery attack}
\label{ATAQUE_NO_MESSAGE} 

Assume now that Eve has the power to replace the tagged message in transit 
between Alice and Bob with a forged tagged message of her own,
$\rho_E$. From this tagged message, Bob will decode the state

\begin{equation}
\rho_\E^E(k)=U^\dagger(k) \rho_E U(k),
\label{LIMBO_1}
\end{equation}

\noindent if the value of the key shared with Alice is
$k$. The probability of Eve being undetected is one when
Eq. (\ref{CONDICION_PROBABILIDAD_UNO}) is satisfied, where
$\rho_\E(k)=U^\dagger(k) \rho_E U(k)$. This condition can be restated as

\begin{equation}
P_i U^\dagger(k) \rho_E U(k) P_i= U^\dagger (k) \rho_E U(k), \quad
\forall k.
\end{equation}

\noindent If we rewrite this equation in terms of the `i-o' decompositions
of its operators, one finds that the $k=0$ case requires $\rho_E$ to be
necessarily of the form

\begin{equation}
\rho_E=\left(
\begin{array}{cc}
\rho_{ii} & 0\\
0 & 0
\end{array}
\right),
\end{equation}

\noindent and that, in the rest of the cases ($\forall k\neq
0$), $U_{io}^\dagger(k)\rho_{ii}=0$, or, equivalently, 

\begin{equation}
\rho_{ii} \in \bigcap_{k\neq 0}{\rm ker}[U_{io}^\dagger(k)].
\end{equation}

In order to make Eve's attack unsuccessful, Alice and Bob should
choose the $U_{io}^\dagger(k)$ such that $\bigcap_{k\neq 0}{\rm
ker}[U_{io}^\dagger(k)]=\{\emptyset\}$. One possibility is
that ${\rm ker}[U_{io}^\dagger(k)]=\{\emptyset\}$, $\forall
k\neq 0$.  This can be accomplished if the dimensions of the spaces $\C$
and $\CP$, $C$ and $D$, respectively, are such that $C\leq D$, since,
in this case, Alice and Bob can always make the rank of $
U_{io}^\dagger(k)$ equal to $C$, $\forall k\neq 0$. Geometrically, this
condition says that no state in $\C$ has null projection over $\CP$
after it has been transformed by $U^\dagger(k)$. In other words, 
the subspace $\C_k$  is different from $\C$, $\forall k>0$.

\section{Discussion}

In the preceding sections we have shown the feasibility of quantum
authentication schemes based on unitary coding sets.  We have given the
conditions that a particular family of unitary operations $\{U(k)\}$,
$k=0, \cdots, K-1$, has to fulfill in order to make, under unitary and
forgery attacks, Eve's probability of success less than one. In the
case of the unitary attack we have shown that ${\rm log}(q+1)$ bits of
classical key are enough when the ratio between the dimensions of the
invalid and valid tag subspaces is $q$, independently of the number
qubits of the original message. This result may seem surprising, but
not too practical since one would expect the effectiveness of such a
protocol to be very low. In fact, to quantitatively study the security
of an authentication protocol based on this scheme, one should find
an appropriate family $\{ U(k) \}$ and obtain lower bounds for $P^f_e$
and $P^u_e$, the error probabilities under forgery and unitary attacks,
respectively.

In the case of the forgery attack,

\begin{equation}
P^f_e= {1 \over K} \sum_{k=0}^{K-1} \mbox{tr}_{\E} \left[P_i U(k)^\dagger
\rho_E U(k) \right] =  {1 \over K} \sum_{k=0}^{K-1} \mbox{tr}_{\E}
\left[P_i(k) \rho_E \right],
\label{EXITO_FORGERY}
\end{equation}

\noindent with $P_i(k)=U(k)P_iU^\dagger(k)$ the projector over $\C_k$,
and $\rho_E$ any state in $\E$ selected by Eve. The above probability
is bounded by the maximum eigenvalue of the hermitian operator
$\sum_{k=0}^{K-1} P_i(k)$, so Alice and Bob's goal is to minimize the
maximum eigenvalue of this operator. For simplicity, assume the dimension
of the total Hilbert space $\E$ is an integer multiple of the dimension of
the code subspace $\C$, $E=pC$, $p>1$. If Alice and Bob choose a family
$\{ U(k) \}$ with  $K \leq p$, such that the projectors $\{ P_i(k) \}$
are mutually orthogonal, then the maximum eigenvalue of $\sum_{k=0}^{K-1}
P_i(k)$ is minimized to one, and $P^f_e \leq 1/K$.

But unfortunately this family of encoding operators is clearly insecure
against the unitary attack. The complete expression for $P^u_e$, the
probability of Eve being unnoticed when Alice prepared $\rho_\E$ under
the unitary attack is:

\begin{equation} 
P^u_e= {1\over K} \sum_{k=0}^{K-1} \mbox{tr}_{\E}
\left[ P_i U^\dagger(k)F_\E \rho_\E(k) F_\E^\dagger U(k) \right]
= {1\over K} \sum_{k=0}^{K-1} \mbox{tr}_{\E} \left[ P_i(k) F_\E
\rho_\E(k) F_\E^\dagger \right],
\label{EXITO_UNITARY} 
\end{equation}

\noindent with $\rho_\E(k)=U(k)\rho_\E U^\dagger(k)$.  Because the
subspaces $\C_k$ obtained after transforming $\C$ with the $U(k)$
above are all orthogonal, disjoint subspaces of $\E$, if Eve chooses a
unitary operator $F_\E$ acting inside each subspace separately with the
general form:

\begin{equation}
F_\E= \sum_{l=0}^{K-1}P_i(l)
F_\E P_i(l), 
\label{SUMA_DIRECTA} 
\end{equation}

\noindent then its action on $\rho_\E(k)$ gives $F_\E \rho_\E(k)
F_\E^\dagger = P_i(k)F_\E \rho_\E(k) F_\E^\dagger P_i(k)$, which is
invariant under projection over $P_i(k)$. Therefore,

\begin{equation} 
P^u_e= {1\over K} \sum_{k=0}^{K-1}
\mbox{tr}_{\E} \left[  F_\E \rho_\E(k) F_\E^\dagger \right] = 1.
\end{equation}

Finding a family $\{ U(k) \}$ producing an acceptable bound for $P^u_e$ is
not easy. One possible family, using the formalism of quantum stabilizer
error correcting codes, has been given by Barnum \cite{BARNUM_2002} with
$P_e \sim (2 + 2m/t)/(2^t + 1)$ for a key of length $2m + O(t)$ bits.
Intuitively, the nature of the conditions on the $\{ U(k) \}$ given in
the preceding Section seems to indicate that $P^u_e$ would decrease with
the overlapping between the subspaces $\C_k$; however, that overlapping
cannot be too big, because if, in the limiting case, the subspaces
$\C_k$ are actually the same subspace, then the probability of failure
under both types of attacks would be one. Specifically, in the forgery
case, $U(k)P_iU^\dagger(k)= P_i$ $\forall k$, and therefore $P^f_e =
\frac{1}{K}\sum_{k=0}^{K-1} \mbox{tr}_{\E} \left(P_i \rho_E \right) =
1$ for any $\rho_E \in \C$. In the case of the unitary attack, if Eve
chooses $F_\E$ such that $[F_\E, P_i ] =0$ then,

\begin{equation}
P^u_e={1\over K}\sum_{k=0}^{K-1} \mbox{tr}_{\E} \left[P_i F_\E \rho_\E
(k) F_\E^\dagger \right] = {1\over K}\sum_{k=0}^{K-1} \mbox{tr}_{\E}
\left[P_i  \rho_\E (k)  \right] =1,
\end{equation}

\noindent where the last equality is obtained using the fact that if
$\rho_E (k) \in \C$, then $P_i  \rho_\E (k)= \rho_\E (k)$, $\forall k$.
Thus we would be looking for an intermediate situation, in which the
$\C_k$ are not  orthogonal neither coincident $C$-dimensional subspaces
inside $\E$.  But what family of $\{ U(k) \}$ is optimal remains an open
problem. It might even be the case that no optimal strategy to handle at
the same time the two types of attack can be found:  Loosely speaking,
if we represent $P^u_e$ and $P^f_e$ against some type of measure of the
overlapping between the subspaces $\C_k$, then one would expect $P^f_e$
to increase from its minimus value to one as the overlapping increases,
and $P^u_e$ to have at least one minimus in an intermediate point between
zero and total overlapping. But the two curves might not cross, or cross
in different ways, making the decision on which is the optimal family
of encoding operators not obvious.

\section{Conclusion}

We have addressed the problem of how to authenticate quantum messages
between two partners (Alice and Bob) connected by an ideal quantum
communication channel. Any authentication process requires a previous
secret between the communicating partners, and we have assumed that Alice
and Bob share a classical secret key. Our authentication scheme uses
a tagging procedure in the transmitting end, and unitary operations
selected by the key (encoding and decoding rules), on both ends of
the channel. In our feasibility analysis, failure can be caused by an
undesired unitary manipulation of the information in transit between the
partners, or by impersonation (forgery of a fake authentication message).
We have shown the conditions that the sets of encoding and decoding rules
must satisfy to make authentication possible.  These conditions are
better stated in terms of geometrical relations between the subspaces
of valid tagged messages selected by each unitary rule.  Specifically,
to protect against forgery, we have shown that no pair of subspaces has
to be coincident; in fact, the failure probability reaches its minimum,
which is inverse in the key length, when the spaces are all disjoint and
orthogonal. On the other hand, to protect against the unitary attack,
the restrictions on the encoding rules are much more involved.  Briefly,
an intermediate situation, one in which the encoding subspaces are neither
coincident nor disjoint, seems to be the desired setting.  In particular,
a key length of order $D/C$, the ratio between the dimension of the
invalid and valid message subspaces, is enough.

Many open questions related to the quantum authentication schemes analysed
deserve further investigation. First, one would like to find the optimal
family of encoding rules protecting against the unitary or even more
general (those based on TPCP maps) attacks, and to give an explicit
expression for the probability of failure in terms of the encoding rules
employed. This may require a more practical definition of the probability
of failure. For instance, Eve might, with high probability, transform
the original message, but only in a way such that the fidelity between
the original and the transformed state be still very high. It could also
be the case that she could strongly transform, without being noticed,
messages inside a particular subspace of the valid message space, and be
noticed if she transforms messages outside it. All these situations would
have to be considered by Alice and Bob in any practical implementation
of the protocol. Another important practical issue is whether an optimal
family of encoding rules against both types of attacks exists.

\begin{acknowledgments}

This work was partially supported by Xunta de Galicia (Spain, grant No.\
PGIDT00PXI322060PR) and the Spanish Government (grant No.\ TIC2001-3217).

\end{acknowledgments}

\end{document}